\documentclass[aps, 10pt, prd,
               notitlepage, twocolumn, superscriptaddress,
               longbibliography,
               nofootinbib, floatfix]{revtex4-2}

\usepackage{amsmath}
\usepackage{amssymb}
\usepackage{amsfonts}
\usepackage[utf8]{inputenc}
\usepackage[T1]{fontenc}
\usepackage{mathrsfs}
\usepackage{anyfontsize}
\usepackage{microtype}

\usepackage[linktocpage,breaklinks]{hyperref}
\usepackage[usenames,dvipsnames]{xcolor}

\usepackage{tensor}
\usepackage{bm}

\usepackage{graphicx}
\usepackage{epsfig}
\usepackage{epstopdf}

\usepackage{natbib}

\hypersetup{colorlinks=true,
            citecolor=NavyBlue,
            linkcolor=NavyBlue,
            urlcolor=NavyBlue}

\newcommand{\ps}{\,(+)}
\newcommand{\mn}{\,(-)}
\newcommand{\nn}{\nonumber \\}
\newcommand{\dd}{{\rm d}}

\newcommand{\ii}{i}
\newcommand{\lm}{\ell m}
\newcommand{\lmw}{\ell m \omega}
\newcommand{\ddr}[1]{\frac{\dd{#1}}{\dd r}}
\newcommand{\ddx}[1]{\frac{\dd{#1}}{\dd x}}
\newcommand{\cA}{{\cal A}}
\newcommand{\cE}{{\cal E}}
\newcommand{\cL}{{\cal L}}
\newcommand{\cLcrit}{{\cal L}_{\rm crit}}
\newcommand{\pd}{\partial}
\newcommand{\mode}{X^{\, (\pm)}_{\lmw}}
\newcommand{\modeUp}{X^{\, (\pm),\,{\rm up}}_{\lmw}}
\newcommand{\modeIn}{X^{\, (\pm),\,{\rm in}}_{\lmw}}
\newcommand{\modeUpIn}{X^{\, (\pm),\,{\rm up/in}}_{\lmw}}
\newcommand{\supIn}{\, (\pm),\, {\rm in}}
\newcommand{\supOut}{\, (\pm),\, {\rm out}}
\newcommand{\Cin}{C^{\,(\pm),\, {\rm in}}_{\lmw}}
\newcommand{\Cout}{C^{\,(\pm),\, {\rm out}}_{\lmw}}
\newcommand{\CoutZ}{C^{\,(+),\, {\rm out}}_{\lmw}}
\newcommand{\CoutRW}{C^{\,(-),\, {\rm out}}_{\lmw}}
\newcommand{\vect}[1]{\boldsymbol{#1}}

\newcommand{\be}{\begin{equation}}
\newcommand{\ee}{\end{equation}}
\newcommand{\X}{X^{\ast}_{\lm}}
\newcommand{\dYdtheta}{\pd_{\theta}Y^{\ast}_{\lm}}
\newcommand{\dYdphi}{\pd_{\phi}Y^{\ast}_{\lm}}
\newcommand{\ddYddphi}{\pd_{\phi\phi}Y^{\ast}_{\lm}}
\newcommand{\half}{\tfrac{1}{2}}

\newcommand{\aei}{\affiliation{Max Planck Institute for Gravitational Physics (Albert Einstein Institute),
D-14476 Potsdam, Germany}}
\newcommand{\um}{\affiliation{Departamento de F\'isica, Universidad de Murcia, Murcia, E-30100, Spain}}
\newcommand{\utub}{\affiliation{Theoretical Astrophysics, University of T\"ubingen,
Auf der Morgenstelle 10, D-72076 T\"ubingen, Germany}}
\newcommand{\uv}{\affiliation{Department of Physics, University of Virginia, Charlottesville, Virginia 22904, USA}}

\begin{document}

\title{Gravitational radiation from a particle plunging into a Schwarzschild black hole: \\
frequency-domain and semirelativistic analyses}

\begin{abstract}
We revisit the classic problem of gravitational wave emission by a test
particle plunging into a Schwarzschild black hole both in the frequency-domain
Regge-Wheeler-Zerilli formalism and in the semirelativistic approximation.
We use, and generalize, a transformation due to Nakamura, Sasaki, and Shibata
to improve the falloff of the source term of the Zerilli function. The faster
decay improves the numerical convergence of quantities of interest, such
as the energy radiated at spatial infinity through gravitational waves.
As a test of the method, we study the gravitational radiation produced by test particles that
plunge into the black hole with impact parameters close to the threshold for scattering. We
recover and expand upon previous results that were obtained using the Sasaki-Nakamura equation.
In particular, we study the relative contributions to the total energy radiated due to waves
of axial and polar parity, and uncover an universal behavior in the waveforms at late times.
We complement our study with a semirelativistic
analysis of the problem, and we compare the two approaches.
The generalized Nakamura-Sasaki-Shibata transformation presented here
is a simple and practical alternative for the analysis of gravitational-wave
emission by unbound orbits in the Schwarzschild spacetime using the
frequency-domain Regge-Wheeler-Zerilli formalism.
\end{abstract}

\author{Hector O. Silva}    \aei
\author{Giovanni Tambalo}   \aei
\author{Kostas Glampedakis} \um  \utub
\author{Kent Yagi}          \uv

\maketitle

\section{Introduction}

The study of gravitational radiation produced by test particles in black-hole spacetimes
has a long history dating back to the early 1970s, and played a central role in the
early development of the understanding of potential gravitational-wave sources~\cite{Rees:1974iy,Smarr:1979ofa}.
In the framework of black-hole perturbation theory, developed by Regge, Wheeler, and Zerilli~\cite{Regge:1957td,Zerilli:1970se}, the pioneering works on this problem
were done by Zerilli~\cite{Zerilli:1970wzz}, Davies et al.~\cite{Davis:1971gg,Davis:1971pa,Davis:1972ud}, Chung~\cite{Chung:1973fn}, and Ruffini~\cite{Ruffini:1973ky}. These
works assumed particles in unbound trajectories that start from spatial infinity and plunge into a Schwarzschild black hole.

Critical to these calculations is the asymptotic behavior of the source term
that encapsulates how the test particle excites the gravitational perturbations.
As an extreme example, the source term of the Teukolsky equation,
that describes the gravitational perturbations of a Kerr black hole~\cite{Teukolsky:1973ha}, diverges at spatial infinity for unbound geodesics.
In principle, this jeopardizes the calculation of physical quantities of interest, such as gravitational waveforms or the energy carried away to infinity by the waves.
To circumvent this problem, one can either develop a regularization scheme~\cite{Detweiler:1979xr,Tashiro:1981ae,Poisson:1996ya,Campanelli:1997sg}, or rewrite the Teukolsky equation to tame the source's asymptotic behavior. Pursuing the latter
approach, Sasaki and Nakamura~\cite{Sasaki:198185,Sasaki:1981kj,Sasaki:1981sx} found their eponymous equation, widely used in the study of unbound geodesics both in Schwarzschild and Kerr spacetimes; see, e.g., Refs.~\cite{Oohara:1983gq,Oohara:1983xip,Oohara:1984ck} and~\cite{Kojima:1983ua,Kojima:1983wwm,Sasaki:1989ca}, respectively, for early work.
A somewhat similar situation also happens when the perturbations of a Schwarzschild black hole are described in terms of the Cunningham-Price-Moncrief~\cite{Cunningham:1978zfa} and Zerilli-Moncrief~\cite{Moncrief:1974am} functions.
Less dramatically, the source term of the Zerilli equation has a slow falloff at spatial infinity. See Hopper~\cite{Hopper:2017qus} for a detailed discussion.

In a serendipitous event, in the course of a related investigation, we learned of a work by Shibata and Nakamura that presents a simple transformation of the Zerilli function to improve the falloff of the source term of the Zerilli equation, and that was applied to the radial infall case only~\cite{Shibata:1992zs}.\footnote{Reference~\cite{Shibata:1992zs} cites an unpublished work by Nakamura and Sasaski when introducing this transformation, that we will refer to as the Nakamura-Sasaki-Shibata transformation.}
Here we revisit this method and extend its application range to problems involving particles plunging with nonzero angular momentum.
This allows us to reexamine some aspects of the gravitational radiation produced by particles that plunge with angular momenta near the threshold for scattering.
This situation is relevant in the context of understanding ultrarelativistic binary black hole collisions, and that was until now only studied in detail using the Sasaki-Nakamura equation; see Berti et al.~\cite{Berti:2010ce} and references therein.
We complement our study of this problem with a calculation of the energy radiated within the semirelativistic approximation of Ruffini and Sasaki~\cite{Ruffini:1981af}.

This paper is organized as follows.
In Sec.~\ref{sec:geo_motion}, we review the motion of test particles
plunging into a Schwarzschild black hole.
In Sec.~\ref{sec:bhpt_rwz}, we provide a summary of Regge-Wheeler-Zerilli
formalism and identify the main issue we want to resolve.
In Sec.~\ref{sec:asympt_sources}, we review and generalize the method of Ref.~\cite{Shibata:1992zs}.
In Sec.~\ref{sec:results}, we describe our numerical methods and present our numerical results.
In Sec.~\ref{sec:kludge}, we compare our results against an analysis in the semirelativistic approximation.
We summarize our findings in Sec.~\ref{sec:conclusions}.
We use the mostly plus metric signature and use geometrical units with $c = G = 1$,
unless stated otherwise.

\section{Geodesic motion}\label{sec:geo_motion}
We consider a particle of mass $\mu$ in geodesic motion the spacetime of a Schwarzschild
black hole of mass $M$, with $\mu / M \ll 1$.
We use Schwarzschild-Droste coordinates $x^{\mu} = \{ t, r, \theta, \phi \}$
in which the spacetime's line element is
\begingroup
\allowdisplaybreaks
\begin{equation}
    \dd s^2 = - f(r) \, \dd t^2 + f^{-1}(r) \, \dd r^2 + r^2 ( \dd \theta^2 + \sin^2\theta \, \dd\phi^2 )\,,
    \label{eq:line_element}
\end{equation}
\endgroup
where $f = 1 - 2 M /r$ and $r = 2 M$ is the location of the event horizon.

\begin{figure}[b]
\includegraphics{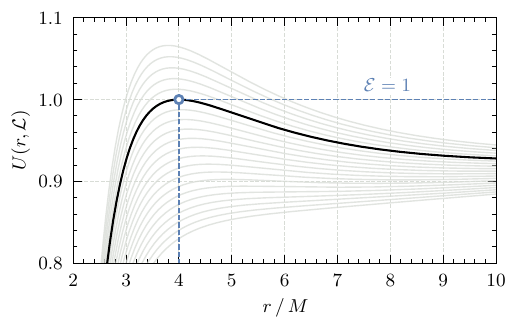}
\caption{
The effective potential $U(r, \cL)$ for various values of $\cL$, which
ranges from $3.25M$ (lowermost solid gray line) to $4.25M$ (uppermost solid
gray line).
For a particle falling from rest from infinity ($\cE = 1$) into the hole, there
is a critical value of $\cLcrit = 4 M$ (solid black line) for which the
particle becomes trapped in a marginally stable circular orbit at $r = 4M$.
This value of $\cL$ separates plunging ($\cL < \cLcrit$) from scattering ($\cL > \cLcrit$) geodesics.
}
\label{fig:eff_potential}
\end{figure}

We assume that the particle starts from rest at infinity with (conserved)
energy $\cE = E / \mu = 1$ and angular momentum $\cL = L / \mu$ per unit mass.
If we chose, without loss of generality, that the particle's motion happens in the
equatorial plane $\theta = \pi / 2$, then we can parametrize the particle's
worldline in terms of the proper time $\tau$ as
$z^{\mu}(\tau) = \{ t_{p}, r_{p}, \pi/2, \phi_p \}$,
and, from the geodesic equation and the timelike constraint $g_{ab} u^{a} u^{b} = -1$,
with $u^{a} = \dd z^{a} / \dd t$, we obtain:
\begin{equation}
    \dot{t}_p = \cE / f_{p}
    \,, \quad
    \dot{\phi}_p = \cL / r_p^2
    \,, \quad
    \dot{r}_p^2 = \cE^2 - U(r_p, \cL) \,,
    \label{eq:geo_ode}
\end{equation}
where $\dot{} = \dd / \dd \tau$ and
\begin{equation}
    U = f \, ( 1 + \cL^2 / r^2) \,,
    \label{eq:eff_potential}
\end{equation}
is the effective radial potential.

The particle's trajectories are classified according to the number of
real roots of $\cE^2 - U$.
For the special case $\cE = 1$ the analysis is simple, and we find the roots to be,
\begin{equation}
    r_{\pm} = \frac{\cL}{4M} \, [ \cL \pm (\cL^2 - 16 M^2)^{1/2} ] \,.
\end{equation}
Hence, a plunging orbit from infinity requires that $\cL < 4 M$, for otherwise the turning
points are real and positive (i.e., the particle is scattered.)
We will define this special value of the angular momentum as
$\cLcrit = 4 M$.

In Fig.~\ref{fig:eff_potential} we show the effective potential~\eqref{eq:eff_potential}
for a range of values $\cL / M \in \{3.25,\, 4.25\}$ (light curves).
The thicker line corresponds to $U(r, \cLcrit)$, which peaks at $r = 4 M$ with value of 1.
Hence, a particle falling from rest and with angular momentum $\cL = \cLcrit$ will be captured in a marginally stable circular orbit.
A particle with $\cL$ larger (smaller) than $\cLcrit$ will scatter (plunge).

\begin{figure}[b]
\includegraphics{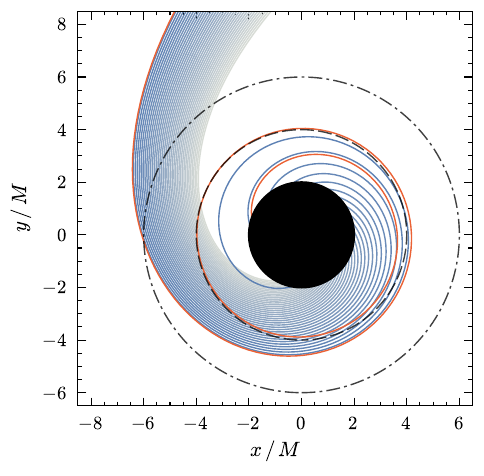}
\caption{
Timelike geodesics with angular momentum per unit mass $\cL
/ M \in \{ 3.25 ,\, 3.9996 \}$ plunging into a Schwarzschild black hole (black
disk) starting from rest at spatial infinity, $\cE = 1$.
The dot-dashed line represents to the innermost stable circular orbit
($r = 6M$) and the dashed line corresponds to the location of the marginally stable circular
orbit ($r = 4 M$).
In the limit $\cL / \cLcrit \to 1$, the particle executes an increasing
fractional number of orbits $\phi / (2 \pi)$, as seen in the red curve that
corresponds to $\cL = 3.9996 M$, or $99.99\%$ of $\cLcrit$.
}.
\label{fig:geodesics}
\end{figure}

It is convenient to rewrite Eq.~\eqref{eq:geo_ode} as first-order in $r$ equations,
\begin{subequations}
\label{eq:geo_ode_radial}
\begin{align}
\dd t_p / \dd r_p   &= - (\cE / f_p) \, (\cE^2 - U_p)^{-1/2} \,,
\\
\dd \phi_p / \dd r_p &= - (\cL / r_p^2) \, (\cE^2 - U_p)^{-1/2} \,,
\end{align}
\end{subequations}
where we have taken $\dot{r}_p < 0$.
We integrate Eqs.~\eqref{eq:geo_ode_radial} with initial conditions
$t_p(r_{\rm max}) = 0$ and $\phi_p(r_{\rm max}) = 0$ at some arbitrarily large $r_p = r_{\rm max}$
down to the horizon, $r_p = 2M$.
In Fig.~\ref{fig:geodesics} we show a sequence of trajectories starting from $\cL / M = 3.25$
and up to $\cL / M = 3.9996$ (i.e., with 99.99\% of $\cLcrit$).
We translate from Schwarzschild-Droste to Cartesian coordinates using
\begin{equation}
x_p = r_p \cos\phi_p \,, \quad
y_p = r_p \sin\phi_p \,, \quad
z_p = 0 \,.
\label{eq:geo_sch_to_cart}
\end{equation}
As the ratio $\cL / \cLcrit$ approaches unity from below, the particle executes an increasing fractional number of orbits $\phi_p / (2 \pi)$,
given approximately by $- (\sqrt{2} \pi)^{-1} \, \log(1 - \cL/\cLcrit)$~\cite{Berti:2009bk}.

\section{Black hole perturbations in the Regge-Wheeler-Zerilli gauge}
\label{sec:bhpt_rwz}

We are interested in calculating the gravitational waves produced
by a particle plunging in a Schwarzschild black hole.
The standard treatment of this problem, in the metric-perturbation formalism, is due
to Regge and Wheeler~\cite{Regge:1957td} and Zerilli~\cite{Zerilli:1970se,Zerilli:1970wzz}.
The problem reduces to solving two equations in the time domain:
\begin{equation}
    \left[ - \frac{\pd^2}{\pd t^2} + \frac{\pd^2}{\pd x^2} - V^{(\pm)}_{\ell}(r) \right] X^{(\pm)}_{\lm}(t,r) = S^{(\pm)}_{\lm}(t,r) \,,
    \label{eq:eqs_rwz_td}
\end{equation}
or, by going to the Fourier domain through
\begin{equation}
    X^{(\pm)}_{\lm}(t,r) = \frac{1}{2\pi} \int_{-\infty}^{\infty} \dd \omega \, e^{- i \omega t} \, X^{(\pm)}_{\lmw}(r)\,,
\end{equation}
we have alternatively
\begin{equation}
    \left[ \frac{\dd^2}{\dd x^2} + \omega^2 - V^{(\pm)}_{\ell}(r) \right] X^{(\pm)}_{\lmw}(r) = S^{(\pm)}_{\lmw}(r) \,.
    \label{eq:eqs_rwz_fd}
\end{equation}
In these equations, $x$ is the tortoise coordinate
\begin{equation}
x = r + 2M \log[r / (2M) - 1]  \,,
\end{equation}
that maps the region $2M < r < \infty$ to $-\infty < x < \infty$.
The superscript $(\pm)$ denotes variables associated
to metric perturbations of polar ($+$) or axial ($-$) parity, and $V^{(\pm)}_{\ell}$ is
an effective potential.
Perturbations of each parity are described by a single master function, known
as the Zerilli $X^{(+)}$ and Regge-Wheeler $X^{(-)}$ functions, respectively.
The effective potentials associated to each of these perturbations bear the same
respective names and are given by
\begin{subequations}
\label{eq:effective_potentials}
\begin{align}
    V_{\ell}^{\ps} &= \frac{f}{r^2 \Lambda^2} \left[
        2 \lambda^2 \left(\Lambda + 1\right)
        + \frac{18 M^2}{r^2} \left( \lambda + \frac{M}{r} \right)
    \right],
    \label{eq:pot_zerilli}
    \\
        V_{\ell}^{\mn} &= \frac{f}{r^2} \left[ \ell(\ell+1) - \frac{6M}{r} \right],
    \label{eq:pot_regge_wheeler}
\end{align}
\end{subequations}
where we defined
\begin{equation}
    \lambda = (\ell + 2) (\ell - 1) / 2
    \quad \textrm{and} \quad
    \Lambda = \lambda + 3M/r \,.
\label{eq:def_lambdas}
\end{equation}
The function $S^{(\pm)}_{\lmw}$ is the source term, responsible for the
excitation of the gravitational perturbations. A detailed derivation of this
source term in the Regge-Wheeler-Zerilli formalism can be found, e.g.,~in Refs.~\cite{Sago:2002fe,Nakano:2003he}, whose
results we quote in Appendix~\ref{app:sources}.

Because the potential $V_{\ell}^{(\pm)}$ vanishes both at the horizon and at infinity,
and provided that the source vanishes sufficiently fast at both boundaries,
the solutions to Eq.~\eqref{eq:eqs_rwz_td} can be written as plane waves for
$x \to \pm \infty$.
We will impose that $\mode$ is purely ingoing at the event horizon
and purely outgoing at spatial infinity, that is,
\begin{align}
    \mode \simeq
    \begin{cases}
        \, \Cin  \, e^{-i \omega x} \quad &x \to -\infty\\
        \, \Cout \, e^{+i \omega x} \quad &x \to +\infty
    \end{cases} \,.
    \label{eq:bcs_inhom}
\end{align}

To solve the inhomogeneous differential equation~\eqref{eq:eqs_rwz_td},
we use the method of variation of parameters.
The method consists of finding two linearly independent solutions, say $\modeIn$
and $\modeUp$, of the homogeneous equations
\begin{equation}
    \left[ \frac{\dd^2}{\dd x^2} + \omega^2 - V^{(\pm)}_{\ell}(r) \right] \modeUpIn = 0 \,.
    \label{eq:eqs_rwz_hom}
\end{equation}
The two solutions differ by the boundary conditions we impose when we solve
Eq.~\eqref{eq:eqs_rwz_hom}, namely,
%
\label{eq:bcs_inup}
\begin{equation}
\modeIn \simeq
    \begin{cases}
        \, e^{-i \omega x}
        \, &x \to -\infty \\
        \, A^{\supIn}_{\lmw} \, e^{- i \omega x} + A^{\supOut}_{\lmw} \, e^{+ i \omega x}
        \, &x \to +\infty
    \end{cases} \nn
\end{equation}
and
\begin{equation}
\modeUp \simeq
    \begin{cases}
        \, B^{\supIn}_{\lmw} \, e^{- i \omega x} + B^{\supOut}_{\lmw} \, e^{+ i \omega x}
        \, &x \to -\infty \\
        \, e^{i \omega x}
        \, &x \to +\infty 
    \end{cases} \,.
    \nonumber
\end{equation}

We use these solutions to define the Wronskian,
\begin{align}
    W^{\,(\pm)}_{\lmw} = \modeIn \, \ddx{\modeUp} - \modeUp \, \ddx{\modeIn} \,,
\label{eq:wronskian}
\end{align}
which is constant in $x$.
We then construct the Green's function as
\begin{align}
    G^{\,(\pm)}_{\lmw}(x,x') &= \frac{1}{W^{\,(\pm)}_{\lmw}}
    \left[ \modeUp(x) \, \modeIn(x') \, \Theta(x-x') \right.
    \nn
    &\quad \left. + \, \modeIn(x) \, \modeUp(x') \, \Theta(x'-x) \right],
    \nn
\label{eq:greens_fun}
\end{align}
where $\Theta(\cdot)$ is the Heaviside step function.
Finally, with the Green's function~\eqref{eq:greens_fun}, we can write the solution of
Eq.~\eqref{eq:eqs_rwz_td} as
\begin{align}
    \mode &= \frac{1}{W^{\,(\pm)}_{\lmw}}
\left[
    \modeUp \int_{-\infty}^{x}  \dd x' \, \modeIn \, S^{(\pm)}_{\lmw}
\right.
\nn
             &\quad + \left.
    \modeIn \int_{x}^{+\infty}  \dd x' \, \modeUp \, S^{(\pm)}_{\lmw}
\right]\,.
\label{eq:gen_sol}
\end{align}
Equation~\eqref{eq:gen_sol} has the desired property of being purely ingoing
at the horizon and purely outgoing at infinity:
\begin{subequations}
\begin{align}
    \lim_{{x \to - \infty}} &\mode \simeq
    \frac{e^{- i \omega x}}{W^{\,(\pm)}_{\lmw}}
    \int_{-\infty}^{\infty}
    \dd x' \, \modeUp \, S^{(\pm)}_{\lmw} \,,
    \label{eq:mode_near_horizon}
    \\
    \lim_{{x \to + \infty}} &\mode \simeq
    \frac{e^{+ i \omega x}}{W^{\,(\pm)}_{\lmw}}
    \int_{-\infty}^{\infty}
    \dd x' \, \modeIn \, S^{(\pm)}_{\lmw} \,,
    \label{eq:mode_near_infty}
\end{align}
\end{subequations}
where we used the boundary conditions on $\modeUpIn$.

We can compare Eq.~\eqref{eq:mode_near_infty} with the boundary conditions
Eq.~\eqref{eq:bcs_inhom} imposed on $\mode$, and make the identification,
\begin{align}
    \Cout = \frac{1}{W^{\,(\pm)}_{\lmw}}
    \int_{-\infty}^{+\infty}
    \dd x' \, \modeIn \, S^{(\pm)}_{\lmw} \,.
\label{eq:out_coef_tor}
\end{align}
This is the main quantity we must numerically calculate to obtain, e.g.,
the time-domain gravitational waveform or the energy radiated to infinity.
More specifically, we will be interested in the spectral energy distribution
in each $(\ell, m)$ mode
\begin{align}
    \frac{\dd E_{\lm}}{\dd \omega} =
    \frac{\omega^2}{64 \pi^2}
    \frac{(\ell+2)!}{(\ell-2)!}
    \left[
    \vert \CoutZ \vert^2
    +
    \frac{4}{\omega^2} \vert \CoutRW \vert^2
    \right],
    \nn
    \label{eq:dEdw}
\end{align}
and in the time-domain Regge-Wheeler and Zerilli mode functions
\begin{equation}
X_{\lm}^{\,(\pm)}(t,x) =
\frac{1}{2\pi} \int_{-\infty}^{+\infty}
\dd \omega \,
\Cout \,
e^{- i \omega (t - x)} \,.
\label{eq:X_td}
\end{equation}
In practice, we note that as $x \to \infty$, the Wronskian becomes
\begin{equation}
W^{\,(\pm)}_{\lmw} = 2 i \omega A^{\supIn}_{\lmw} \,,
\label{eq:wronskian_infty}
\end{equation}
and that we can also rewrite Eq.~\eqref{eq:out_coef_tor} as
\begin{align}
    \Cout = \frac{1}{W^{\,(\pm)}_{\lmw}}
    \int_{2M}^{+\infty}
    \frac{\dd r'}{f} \, \modeIn \, S^{(\pm)}_{\lmw} \,.
\label{eq:out_coef_sch}
\end{align}
Because $f \simeq 1$ and $\modeIn \simeq \exp(\pm i \omega x)$ as $x \to \infty$, we see that
the convergence of this integral depends critically on the asymptotic properties of
$S^{(\pm)}_{\lmw}$.

\section{Asymptotic behavior of the frequency domain sources}
\label{sec:asympt_sources}

We now review the properties of $S_{\lmw}^{\ps}$ in two cases
of interest.
We will start by reviewing the case in which the particle falls radially, $\cL
= 0$, into the black hole and how Ref.~\cite{Shibata:1992zs}
improved the asymptotic behavior of $S_{\lmw}^{\ps}$.
We will then show one way in which the Nakamura-Sasaki-Shibata transformation can
be generalized to general plunging trajectories, ${\cal L} \neq 0$, and
discuss the asymptotic properties of this new source term.

\subsection{The case of radial infall}

\begin{figure*}[t]
\includegraphics{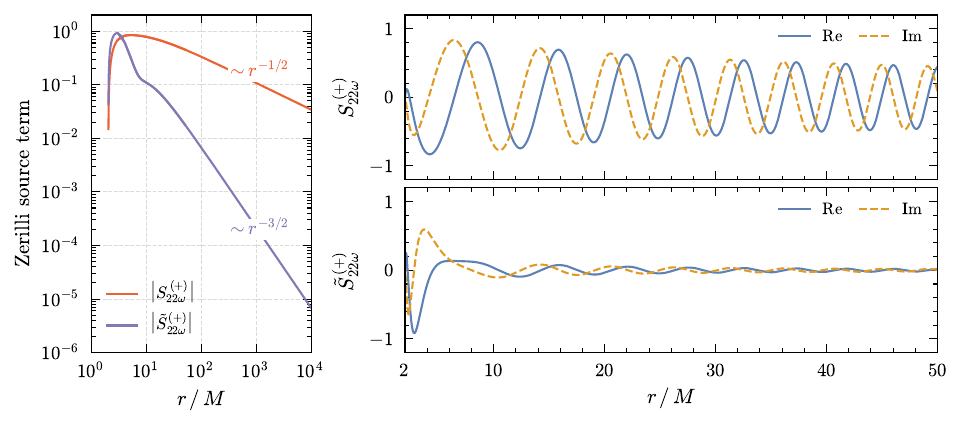}
\caption{The $\ell = m = 2$ source term for the Zerilli equation for a radially plunging particle starting from
rest at spatial infinity, for $M \omega = 0.3$.
Left panel: the absolute values of the original Zerilli source term, $S^{\ps}$, and its Nakamura-Sasaki-Shibata-transformed version, $\tilde{S}^{\ps}$. Right panels: the real (``Re'') and imaginary (``Im'') parts of $S^{\ps}$ (top) and $\tilde{S}^{\ps}$ (bottom).
Both sources vanish at the event horizon $r = 2M$, and the faster asymptotic decay of $\tilde{S}^{\ps}$ improves the convergence of Eq.~\eqref{eq:out_coef_tor}.
The source terms are largest near $r \simeq 3M$, which corresponds approximately to the
location of the peak of the Zerilli potential $r^{(+)}_{\ell=2,\,{\rm peak}} \simeq 3.1 M$
and of the light ring $r=3M$.
}
\label{fig:zerilli_sources_L0}
\end{figure*}

When $\cL = 0$, $S^{\mn}_{\lmw}$ vanishes and $S^{\ps}_{\lmw}$ simplifies to~\cite{Davis:1971pa,Davis:1971gg}
\begin{align}
S^{\ps}_{\lmw} &=
- \, 8 \pi \mu \cA_{\lm}
\frac{f}{ r \Lambda}
\left[
\sqrt{\frac{r}{2M}}
- \frac{2 i}{\omega} \frac{\lambda}{r \Lambda}
\right]
e^{i \omega t_{p}(r)} \,,
\label{eq:source_z_radial_davis}
\end{align}
where ${\cal A}_{\lm} = Y^{\ast}_{\lm}(\pi/2, \phi) \exp(im\phi)$, $Y_{\lm}$ are the spherical harmonics, the asterisk indicates complex conjugation,
and $t_{p}$ is given by
\begin{align}
    \frac{t_p}{2M} =
    - \frac{2}{3} \left( \frac{r}{2M} \right)^{\tfrac{3}{2}}
    - 2 \left( \frac{r}{2M} \right)^{\tfrac{1}{2}}
    + \log\left[ \frac{\sqrt{r/(2M)} + 1}{ \sqrt{r/(2M)} - 1 } \right].
    \nn
\end{align}

We find that the near-horizon and spatial-infinity behaviors of
$S^{\ps}_{\lmw}$ are
\begin{align}
    S^{\ps}_{\lmw} \simeq
    \begin{cases}
        0        &\quad x \to -\infty \\
        x^{-1/2} &\quad x \to +\infty \\
    \end{cases} \,.
    \label{eq:source_z_asympt}
\end{align}
Hence the integral Eq.~\eqref{eq:out_coef_tor} converges slowly
at spatial infinity.
To improve the convergence, Ref.~\cite{Shibata:1992zs} proposed
the substitution\footnote{Our notation differs from that used in Ref.~\cite{Shibata:1992zs}.
See also Ref.~\cite{Tashiro:1981ae}, for a similar substitution in the context of the Teukolsky equation.}
\begin{equation}
    X^{\ps}_{\lmw} = \tilde{X}^{\ps}_{\lmw} + Q_{\lmw} \,,
    \label{eq:sn_trans}
\end{equation}
where
\begin{equation}
Q_{\ell m \omega} =
-\frac{8 \pi \mu \cA_{\lm}}{\omega^2}
\frac{f}{r \Lambda}
\sqrt{\frac{2M}{r}}
e^{\ii \omega t_{p}} \,.
\label{eq:sn_qfun}
\end{equation}
The function $Q_{\lmw}$ vanishes at the event horizon $x \to -\infty$ and decays as $x^{-3/2}$ for $x \to \infty$.
Thus, $X^{\ps}_{\lmw} \simeq \tilde{X}^{\ps}_{\lmw}$ in the latter limit.

We then insert Eq.~\eqref{eq:sn_trans} in  Eq.~\eqref{eq:eqs_rwz_td} to find
\begin{equation}
    \left[ \frac{\dd^2}{\dd x^2} + \omega^2 - V^{\ps}_{\ell} \right]
    \tilde{X}^{\ps}_{\lmw} = \tilde{S}^{\ps}_{\lmw} \,,
    \label{eq:eq_zerilli_sn}
\end{equation}
where $\tilde{S}^{\ps}_{\lmw}$ is a new source term given by
\begin{equation}
    \tilde{S}^{\ps}_{\lmw} =
    S^{\ps}_{\lmw} -
    \left[
    \frac{\dd^2}{\dd x^2}
    + \omega^2
    - V^{\ps}_{\ell}
    \right] Q_{\lmw} \,.
    \label{eq:source_z_radial_sn}
\end{equation}
The asymptotic behaviors of the new source term are
\begin{align}
    \tilde{S}^{\ps}_{\lmw} \simeq
    \begin{cases}
        0        &\quad x \to -\infty \\
        x^{-3/2} &\quad x \to +\infty \\
    \end{cases} \,,
    \label{eq:sn_source_z_asympt}
\end{align}
making the integral in Eq.~\eqref{eq:out_coef_tor} converge faster.
One can verify that it is the second $x$ derivative of $Q_{\lmw}$ in Eq.~\eqref{eq:source_z_radial_sn}
that yields a term that decays as $r^{-1/2}$ as $x \to \infty$, and
that cancels the leading-order asymptotic term in the large-$x$ expansion of
the original source term.
We note that both $X^{\ps}_{\lmw}$ and $\tilde{X}^{\ps}_{\lmw}$ satisfy
the same homogeneous equation.
Hence, we can calculate $\Cout$ with the mere
replacement $S^{\ps}_{\lmw} \to \tilde{S}^{\ps}_{\lmw}$ in Eq.~\eqref{eq:out_coef_sch}.

In Fig.~\ref{fig:zerilli_sources_L0}, we show both the
original~\eqref{eq:source_z_radial_davis} and the
Nakamura-Sasaki-Shibata-transformed~\eqref{eq:source_z_radial_sn} source terms of the
Zerilli equation for a particle falling radially into the black hole,
and $\ell = m = 2$ and $M \omega = 0.3$.
In the left panel we show their absolute values. In the right panels we show
the real (solid line) and imaginary (dashed line) parts of the original (top
panel) and new (bottom panel) source terms.
The faster decay of Eq.~\eqref{eq:source_z_radial_sn} accelerates the
convergence of Eq.~\eqref{eq:out_coef_tor}.
The value $M \omega = 0.3$
is approximately the real part of the fundamental quasinormal mode frequency
of a Schwarzschild black hole~\cite{Chandrasekhar:1975zza} and dominates the emission of energy in
the form of gravitational waves~\cite{Detweiler:1979xr}.
The source terms are largest near $r \simeq 3M$, which corresponds approximately to the
location of peak of the Zerilli potential $r^{(+)}_{\ell=2,\,{\rm peak}} \simeq 3.1 M$,
and of the light ring $r=3M$.
%

The success of the Nakamura-Sasaki-Shibata transformation opens two questions. First,
can we, even for a radially infalling particle, make the source term decay
more rapidly? Second, can we generalize Eqs.~\eqref{eq:sn_trans}
and~\eqref{eq:sn_qfun} to the case in which the particle plunges with a nonzero angular momentum?
In the next section we give positive answers to both questions.

\subsection{The case of general plunging trajectories}\label{subsec:general_case_plunge}

We seek a minimal extension of Eqs.~\eqref{eq:sn_trans}
and~\eqref{eq:sn_qfun} for the case of a plunging particle with general
angular momentum $\cL \neq 0$.
We focus only on the source term in the Zerilli equation, because the source term in the Regge-Wheeler
equation already has a fast falloff:
\begin{equation}
    S^{\mn}_{\lmw} \simeq x^{-2}\,,\quad x \to \infty \,.
\end{equation}
We would like to make the source term in the Zerilli equation to decay at
least as rapidly as the source term for the Regge-Wheeler equation.

\begin{figure*}[t]
\includegraphics{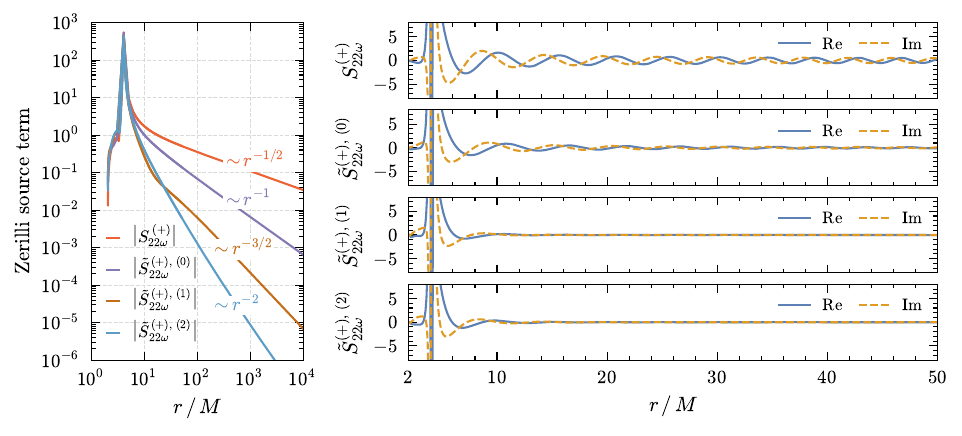}
\caption{The $(\ell, m) = (2, 2)$ source term for the Zerilli equation for a plunging particle with $\cL = 0.9999 \, \cL_{\rm crit}$ starting from rest at spatial infinity, for $M \omega = 0.3$.
Left panel: the absolute values of the original Zerilli source term, $S^{\ps}$, and its Nakamura-Sasaki-Shibata-transformed version, $\tilde{S}^{\ps,\,(N)}$, for $N = 0$, 1, and 2  in the expansion~\eqref{eq:gen_sn_qfun}.
Right panels: the real (``Re'') and imaginary (``Im'') parts of $S^{\ps}$ (top) and $\tilde{S}^{\ps,\,(N)}$ (three
remaining panels) for increasing values of $N$.
The source term is largest near $r \simeq 4 M$, that corresponds to the location of the marginally stable circular orbit
at which the particle particle executes an increasing number of revolutions in the limit $\cL / \cLcrit \to 1$.
}
\label{fig:zerilli_sources_Lneq0}
\end{figure*}

Our strategy is to retain Eq.~\eqref{eq:sn_trans}, but generalize
$Q_{\lmw}$ to the form
\begin{align}
Q_{\ell m \omega} &= -
\frac{8 \pi \mu\cA_{\lm}}{\omega^2}
\frac{f}{\Lambda r}
\left[
    \, \sum_{n=0}^{N} a_{n} \, \left({2M}/{r}\right)^{{(n+1)}/{2}}  \,
\right]
\nn
&\quad \times e^{\ii (\omega t_{p} - m \phi_{p})} \,.
\label{eq:gen_sn_qfun}
\end{align}
Proceeding as in the previous section, we arrive at the same Eq.~\eqref{eq:source_z_radial_sn},
where $S^{\ps}_{\lmw}$ is now given by a more complicated formula,
whose form can be found in Eq.~\eqref{eq:source_z}, Appendix~\ref{app:sources}.
We then expand Eq.~\eqref{eq:source_z_radial_sn} in a power series
in $r$ for $r \to \infty$, and collect powers in $r$.
Next, we fix the coefficients $a_n$ in such a way to cancel each successive
power of $r$ in this power series. This yields
\begin{equation}
    a_{0} = 1\,, \quad
    a_{1} = \frac{i m}{\ell(\ell+1)} \frac{\cL}{M}\,, \quad
    a_{2} = 1 - \frac{\cL^2}{8M^2}\,,
    \label{eq:gen_sn_coefs}
\end{equation}
by working up to $N=2$. Higher-order terms can be obtained by following the same procedure
just outlined.
We notice there is no dependence on $\cL$ at order $N = 0$ and that
we recover Eq.~\eqref{eq:sn_qfun} in this case. The next correction
to the radial infall case occurs at $n=2$.

In Fig.~\ref{fig:zerilli_sources_Lneq0} we show the original source term of the Zerilli equation
$S^{\ps}_{\lmw}$ and its transformed version $\tilde{S}^{\ps, \, (N)}_{\lmw}$ (for $N =$ 0, 1, and 2)
for $\ell = m = 2$, $M \omega = 0.3$ and a near-critical geodesic with $\cL = 0.9999 \, \cL_{\rm crit}$.
We chose this value of $\cL$ as a ``stress test'' for the method, due to the large amplitude of the source term around the location of the
marginally stable circular orbit, $r = 4M$. This peak dominates over the
peak at light ring $r = 3M$ present in Fig.~\ref{fig:zerilli_sources_L0}.
The left panel shows the absolute values of the various source terms, whereas in the right panels
we show their real (solid lines) and imaginary parts (dashed lines).

We see that the original Nakamura-Sasaki-Shibata transformation ($N=0$) results in a
slower decaying source term, $r^{-1}$,  when applied to case in which $\cL \neq 0$,
when compared to the radial infall case, $r^{-3/2}$.
This can be understood by examining the asymptotic behavior of $S^{\ps}_{\lmw}$:
\begin{align}
    S^{\ps}_{\lmw}
    &= - \frac{4 \pi \mu \cA_{\lm}}{ \lambda M}
    \left[
        \sqrt{\frac{2M}{r}}
        + \frac{im}{\ell(\ell+1)} \frac{\cL}{M} \frac{2M}{r}
    \right.
    \nn
    &\left.
        + \left(\frac{\cL^2}{8M^2} + \frac{3 + 2 \lambda}{2 \lambda} \right) \left( \frac{2M}{r} \right)^{3/2}
    \right] + {\cal O}(r^{-2})\,.
    \label{eq:src_z_inf}
\end{align}
This expansion shows that the first term containing the angular momentum $\cL$
appears at order $r^{-1}$, and it cannot be canceled with
Eq.~\eqref{eq:sn_qfun}.

\section{Numerical methods and results}
\label{sec:results}

In this section we describe the numerical methods we use to compute Eq.~\eqref{eq:out_coef_sch},
and show a few applications of solving this equation using the generalized Nakamura-Sasaki-Shibata
transformation~\eqref{eq:gen_sn_qfun}.

\subsection{Numerical methods}

To evaluate our main quantity of interest, namely the amplitude $C^{(\pm),\,\rm out}_{\lmw}$ we proceed as follows.
\begin{enumerate}
    \item Choose a value of $\ell$, $|m| \leq  \ell$, and $M \omega$.
    \label{step:lmw}
    \item Choose a value of the angular momentum $\cL$, and solve the geodesic
    equations~\eqref{eq:geo_ode_radial}, to obtain $t_{p}(r)$ and $\phi_{p}(r)$. We use initial conditions
    $t_{p}(r_{\rm min}) = 0$ and $\phi_p(r_{\rm min}) = 0$, and integrate from $r_{\rm min} = 2 \, (1 + 10^{-4}) M$
    up to $r_{\rm max} = 500 / \omega$.
    \label{step:geodesic}
    \item If $\ell + m$ is even, integrate the homogeneous Zerilli equation.
    If $\ell + m$ is odd, integrate the homogeneous Regge-Wheeler equation; see
    Eqs.~\eqref{eq:eqs_rwz_hom} and~\eqref{eq:effective_potentials}.
    \label{step:z_or_rw}
    \item Integrate the equation from the previous step with ``in'' boundary
    conditions, from $r_{\rm min} = 2 \, (1 + 10^{-4}) M$ to $r = r_{\rm max}$.
    From $X^{(\pm)}_{\lmw}$ and $\dd X^{(\pm)}_{\lmw} / \dd r$ at $r_{\rm max}$
    calculate $A^{(\pm), \, {\rm in},\, {\rm out}}_{\lmw}$. See Appendix~\ref{app:how_to_find_wronskian}.
    \item Calculate the Wronskian~\eqref{eq:wronskian_infty} and the source $S^{(\pm)}_{\lmw}$
    to evaluate the integral in Eq.~\eqref{eq:out_coef_sch}.
    \label{step:get_cout}
    \item Repeat steps~\ref{step:lmw} through~\ref{step:get_cout}, scanning the range
    $M \omega  \in [5 \times 10^{-3}, \, 1.5]$ in steps of size $\Delta (M \omega) = 5 \times 10^{-3}$,
    for $\ell = 2$ to 6, covering all $|m| \leq \ell$ in between.
    \label{step:loop}
\end{enumerate}

Our code is written in Mathematica.
In step~\ref{step:z_or_rw}, we validated our integration of the homogeneous equations~\eqref{eq:eqs_rwz_hom}
by comparison against the integrators available in the Black Hole Perturbation Toolkit~\cite{BHPToolkit}.
Recursion relations for the near-horizon and far-field expansions
of the Regge-Wheeler and Zerilli functions are summarized in Appendix~\ref{app:up_sol}.
In step~\ref{step:get_cout}, we calculate the Wronskian
as explained in Appendix~\ref{app:how_to_find_wronskian},
and it is useful to rewrite
Eq.~\eqref{eq:out_coef_sch} as a differential equation~\cite{Warburton:2013lea,Leather:2023dzj}:
\begin{equation}
\frac{\dd \Cout}{\dd r} = \frac{1}{W^{\,(\pm)}_{\lmw}} \,
f^{-1} \, \modeIn \, {S}^{\,(\pm)}_{\lmw} \,.
\label{eq:out_coef_ode}
\end{equation}
We integrate Eq.~\eqref{eq:out_coef_ode}, with initial condition $C^{\,(\pm),\, \rm out}_{\lmw}(r_{\rm min}) = 0$ up to $r_{\rm max}$,
using the Regge-Wheeler source~\eqref{eq:source_rw} or the Nakamura-Sasaki-Shibata-transformed
Zerilli source~\eqref{eq:source_z_radial_sn}, depending on whether $\ell + m$ is odd or even, respectively.
In the context of Eq.~\eqref{eq:out_coef_ode}, the new Zerilli source term reduces the value of $r_{\rm max}$ at which ${\dd \Cout}/{\dd r}$ becomes zero to the desired accuracy.
We decided here only, with respect to the rest this work, to integrate the geodesic equations
from near the horizon outwards. In particular, when integrating from spatial infinity, $t_p$ acquires a large value around $r \approx 2M$. This causes the factor $\exp(i \omega t_p)$, which appears in the source, to become highly oscillatory, and thus sensitive to our choice of $r_{\rm min}$ and $\omega$.
Consequently, the phase (but not the amplitude) of the $\Cout$ failed to converge in our numerical calculations. We solved this issue by fixing the initial conditions at the horizon instead of spatial infinity.
At last, we performed steps~\ref{step:lmw} to \ref{step:loop} for the range of
angular momentum values $\cL =$~$\{ 0$, 0.25, 0.5, 0.75, 0.9, 0.99, 0.999, $0.9999\}\cLcrit$.

\subsection{Energy spectrum}

\begin{figure*}[bht!]
\includegraphics{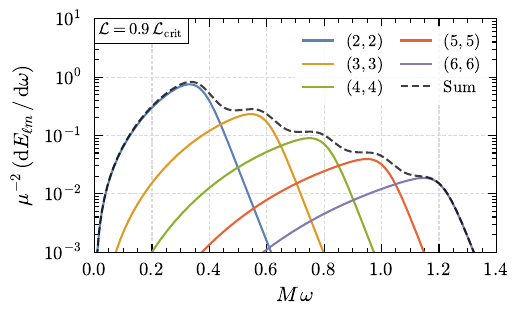}
\includegraphics{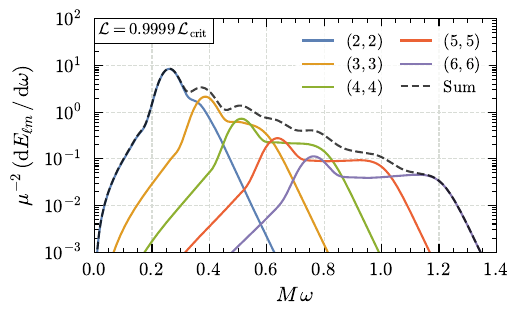}
\\
\includegraphics{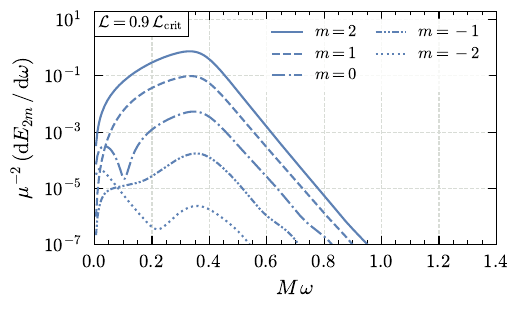}
\includegraphics{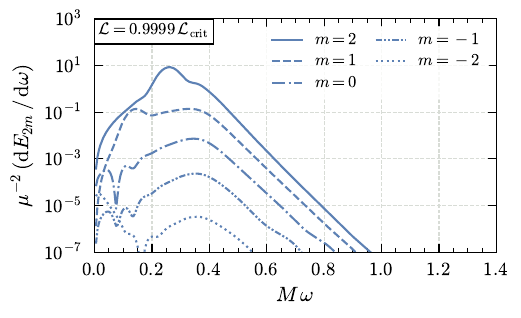}
\caption{The energy spectra for a particle plunging with angular momentum $\cL = 0.9 \, \cL_{\rm crit}$ (left panels) and $\cL = 0.9999 \, \cL_{\rm crit}$ (right panels) into a Schwarzschild black hole.
Top panels: the energy spectra from the multipoles $\ell = m$ from 2 to 6, (colored lines), and their sum (dashed line).
Bottom panels: the energy spectra for the quadrupole perturbation $\ell = 2$ and all $|m| \leq 2$ in between.
Note the different ranges in the ordinates across the panels.
Our results agree with Ref.~\cite{Berti:2010ce}, which solved the Sasaki-Nakamura equation instead.}
\label{fig:dEdw_bhpt}
\end{figure*}

We first consider the energy spectrum for the particle in the near-critical limit $\cL \approx \cLcrit$.
To our knowledge, this situation was studied only in the frequency domain using the Sasaki-Nakamura equation~\cite{Berti:2010ce} or using the
Zerilli-Moncrief~\cite{Moncrief:1974am} and Cunningham-Price-Moncrief~\cite{Cunningham:1978zfa} master functions in Ref.~\cite{Hopper:2017iyq}.
To validate our calculations, we focus on two cases, $\cL / \cL_{\rm crit} = 0.9$ and $0.9999$, which were examined in detail by Berti et al.~\cite{Berti:2010ce}.

In Fig.~\ref{fig:dEdw_bhpt}, we show the energy spectrum~\eqref{eq:dEdw} for $\cL = 0.9 \, \cLcrit$ (left panels)
and $\cL = 0.9999 \, \cLcrit$ (right panels).
The top panels show the spectral energy for multipoles $\ell = m$ ranging from 2 to 6
(different line colors) and their sum (dashed line).
The bottom panels show the quadrupole mode $\ell = 2$ and all azimuthal
contributions $|m| \leq 2$ (different line styles.)
Our results are in excellent agreement with those of Ref.~\cite{Berti:2010ce}; cf.~Figs.~10 and~11 therein.

For $\cL = 0.9 \, \cLcrit$, the energy radiated in each multipole $\ell=m$ has a single maximum.
The location $M \omega$ of these peaks coincide approximately with the real part of the fundamental ($n=0$) quasinormal mode frequency
of a Schwarzschild black hole $M \omega_{\ell m 0}$, as first observed by Detweiler and Szedenits~\cite{Detweiler:1979xr}.
For reference, these values are ${\rm Re}[M \omega_{\ell m 0}] \simeq \{ 0.373,\, 0.599, \, 0.809,\, 1.012,\, 1.212\}$, for $\ell \in \{2,\, 6\}$~\cite{Chandrasekhar:1985kt,Leaver:1985ax,BertiTab}.
(Due to the spherical symmetry of the Schwarzschild solution, all quasinormal modes with $|m| \leq \ell$, at fixed $\ell \geq 2$,
are degenerate.)

In the near-critical limit $\cL = 0.9999~\cLcrit$, the spectral energy distribution has a peak at $M \omega < {\rm Re}[M \omega_{\ell m 0}]$.
This peak corresponds to $m$ times the particle's orbital frequency $M \Omega_{\rm orb} = 1/8$ at the marginally stable circular orbit at $r = 4 M$ (cf.~Fig.~\ref{fig:geodesics}), that is
\begin{equation}
    M \omega = m \, M \, \Omega_{\rm orb} = m/8 \,.
    \label{eq:peak_geo}
\end{equation}
Therefore, the energy emitted at these frequencies is dominated by the particle's geodesic motion. This is not surprising given that we saw that the maximum amplitude of the source term of the Zerilli equation is located at $r \simeq 4 M$,
as we discussed in Fig.~\ref{fig:zerilli_sources_Lneq0}.
For moderate values of $\ell$, the contributions to the energy driven by the particle's orbital motion and the quasinormal mode excitation overlap, while as $\ell \gg 1$ their contributions separate sufficiently to result in two peaks in the energy spectra; see the $\ell = m = 5$ and~6 in Fig.~\ref{fig:dEdw_bhpt}.
For $\ell = m = 5$ the peaks are located at $M \omega \simeq 0.637$ and $0.919$, while for
$\ell = m = 6$ they are found at $M \omega \simeq 0.763$ and $1.125$.
We can estimate this separation as follows.
First, we use the geometrical-optics (eikonal) limit~\cite{Press:1971wr,Goebel:1972ApJ,Cardoso:2008bp}, to approximate the real part
of the quasinormal mode frequency as
\begin{equation}
{\rm Re}[M \omega_{\rm eik}] \simeq \ell \, \Omega_{\rm LR} = \ell / (3 \sqrt{3}),
\label{eq:real_eik_qnm}
\end{equation}
where $M \Omega_{\rm LR} = 1/(3 \sqrt{3})$ is the orbital frequency of a null geodesic
at the light ring. Equation~\eqref{eq:real_eik_qnm} can also be obtained from the $\ell \gg 1$
limit of a Wentzel–Kramers–Brillouin approximation to the calculation of black-hole quasinormal modes~\cite{Schutz:1985km,Iyer:1986np}.
We can then estimate the separation between the ``quasinormal-mode'' and ``geodesic'' peaks
by taking the difference of Eqs.~\eqref{eq:peak_geo} and~\eqref{eq:real_eik_qnm}:
\begin{equation}
    (\Delta M \omega)_{\rm peak} \simeq \frac{1}{3\sqrt{3}}
    \left(
    \ell - \frac{3 \sqrt{3}}{8}m
    \right),
\end{equation}
which is valid for $\ell \gg 1$, $\cL / \cLcrit \simeq 1$, and $\cE = 1$. For $\ell = m = 6$, we find
$(\Delta M \omega)_{\rm peak} \simeq 0.404$, in fair agreement with the
numerical result $\simeq 0.362$, obtained from the difference between the two peak locations.

Although not studied here, it is interesting to analyze the ultrarelativistic limit,
in which the particle plunges with an energy $\cE \to \infty$.
In this limit, the ``geodesic'' peak moves rightwards,
eventually overlapping with the ``quasinormal-mode'' peak; see Ref.~\cite{Berti:2010ce}, Fig.~10.
This occurs because as $\cE \to \infty$, the particle's marginally stable
circular orbit coincides with the light ring, hence $\Omega_{\rm orb} \simeq \Omega_{\rm LR}$~\cite{Berti:2009bk}.
In this limit, we then have
\begin{equation}
    (\Delta M \omega)_{\rm peak} = (\ell - m) \, (3 \sqrt{3}),
\end{equation}
which is valid for $\ell \gg 1$, $\cL / \cLcrit \simeq 1$, and $\cE \to \infty$.
The peak separation vanishes when $\ell = m$, reproducing the results of Ref.~\cite{Berti:2010ce}.

\subsection{Total energy}

From the energy spectrum, we can compute the total energy $\Delta E$ emitted in form of gravitational waves by integrating the spectrum density and summing the contributions
from all multipoles:
\begin{equation}
    \Delta E =
    \sum_{\ell = 2}^{\infty} \, \Delta E_{\ell} =
    \sum_{\ell = 2}^{\infty} \sum_{m = -\ell}^{\ell} \,
    \int_{0}^{+\infty} \dd \omega \,
    \frac{\dd E_{\lm}}{\dd \omega} \,.
    \label{eq:total_energy}
\end{equation}
To understand the relative contribution due to perturbations of each parity, we also define $\Delta E^{\,(\pm)}$,
where the subscript $(+)$ means we add only the contributions to the energy coming from multipoles
for which $\ell + m$ is even and $(-)$ when $\ell + m$ is odd.

For particles that plunge from infinity initially from rest, Oohara and Nakamura~\cite{Oohara:1983xip}, based
on an earlier observation by Davis et al.~\cite{Davis:1971gg} for the case $\cL = 0$, proposed the empirical relation
\begin{equation}
(M / \mu^2) \, \Delta E_{\ell} = a \, \exp(- b \, \ell \,),
\label{eq:on_fit}
\end{equation}
between energy and multipole $\ell$.
We fit Eq.~\eqref{eq:on_fit} to the outcome of computing $\Delta E_{\ell}$, truncating the $\ell$ sum at $\ell_{\rm max} = 6$.
Table~\ref{tab:on_fit} shows the fitting coefficients and the total energy emitted, including the individual polar and axial contributions. The results for the total energy agree with those of Ref.~\cite{Berti:2010ce}, Table II, with less than 1\%  difference, when comparison is possible.
As observed by Oohara and Nakamura, the value of $a$ increases while that of $b$ decreases
as $\cL$ approaches $\cLcrit$. In other words, the fraction of the total energy radiated contained in each $\ell$ pole tends to ``even out'' as $\cL$ increases.

\begin{figure}[htb]
\includegraphics{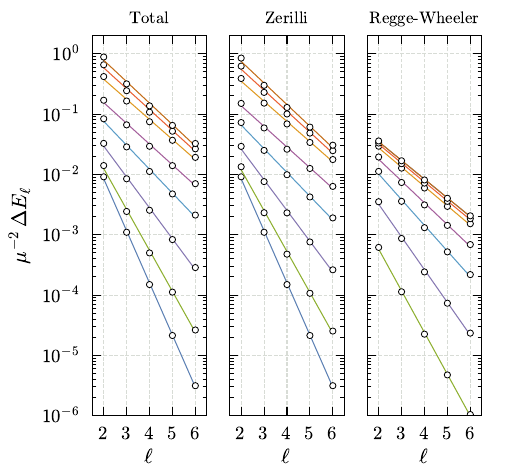}
\caption{Energy radiated to infinity
in the plunge process. From left to right, the panels
show the total energy radiated, only through the parity even (``Zerilli'') perturbations, and
only through the parity odd (``Regge-Wheeler'') perturbations. The markers correspond to Eq.~\eqref{eq:total_energy},
whereas the straight lines corresponds to the fit~\eqref{eq:on_fit} suggested by Oohara and Nakamura~\cite{Oohara:1983xip}. The lines starting from the bottom correspond to $\cL = 0$ and increase
to $\cL = 0.9999 \, \cLcrit$ as one moves upwards.
The energy emitted in the dominant quadrupole mode in the axial-parity radiative channel is
always smaller than that emitted through the polar one, at fixed $\cL$. When $\cL = 0$, there is no energy radiated in the axial-parity channel.
}
\label{fig:energy_total}
\end{figure}

Figure~\ref{fig:energy_total} illustrates these observations. For example, we see in the leftmost panel that $\Delta E_{2}$ varies from being about four orders of magnitude larger than $\Delta E_{6}$, to being some more than ten times larger as $\cL$ approaches criticality.
The individual contributions of the polar and axial perturbations to the energy are shown in the middle and rightmost panels, respectively.
We see that the energy emitted through the ``axial radiative channel'' is always subdominant relative to the polar one, even for near-critical values of $\cL$; the extreme case is, of course, when $\cL = 0$ where $\Delta E^{\,(-)}$ vanishes.

\begin{table}[t]
\begin{tabular}{l c c c c c c c c c}
\hline
\hline
$\cL / \cLcrit$ & $a$ & $b$ & $a^{\,(+)}$ & $b^{\,(+)}$ & $a^{\,(-)}$ & $a^{\,(-)}$ & $\Delta E$ & $\Delta E^{\,(+)}$ & $\Delta E^{\,(-)}$ \\
                &     &     &             &             &             &             & $\cdot 10^{-2}$ &  $\cdot  10^{-2}$ & $\cdot 10^{-4}$ \\
\hline
0.0000 & 0.45 & 1.99 & 0.45 & 1.99 & --   & --   & 1.04 & 1.04 & --   \\ \hline
0.2500 & 0.29 & 1.56 & 0.27 & 1.56 & 0.01 & 1.59 & 1.71 & 1.63 & 7.59 \\ \hline
0.5000 & 0.31 & 1.18 & 0.27 & 1.17 & 0.04 & 1.25 & 4.48 & 4.01 & 47.4 \\ \hline
0.7500 & 0.47 & 0.91 & 0.40 & 0.91 & 0.07 & 0.98 & 13.0 & 11.4 & 169  \\ \hline
0.9000 & 0.76 & 0.79 & 0.66 & 0.79 & 0.09 & 0.83 & 28.7 & 25.4 & 321  \\ \hline
0.9900 & 1.75 & 0.77 & 1.63 & 0.77 & 0.12 & 0.74 & 71.5 & 66.2 & 525  \\ \hline
0.9990 & 2.88 & 0.80 & 2.75 & 0.80 & 0.13 & 0.72 & 109  & 103  & 609  \\ \hline
0.9999 & 4.02 & 0.82 & 3.89 & 0.82 & 0.15 & 0.71 & 144  & 137  & 669  \\ \hline
\hline
\end{tabular}
\caption{Coefficients $a$ and $b$ in the Oohara-Nakamura relation~\eqref{eq:on_fit} and
the total energy $(\mu^2 / M ) \, \Delta E$ for different values of angular momentum $\cL$.
The superscripts refer to polar $(+)$ and axial-parity $(-)$ perturbations.}
\label{tab:on_fit}
\end{table}

\begin{figure}[t]
\includegraphics{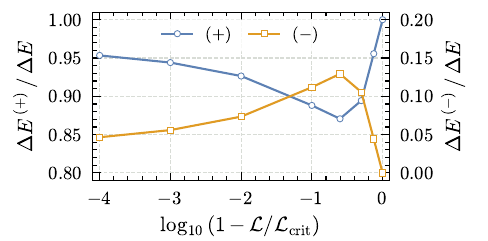}
\caption{Fraction of the total energy radiated by the polar and axial perturbations in the plunge process. We show
both the polar [$(+)$, left ordinate] and axial [$(-)$, right ordinate] total energies as functions of the particle's angular momentum. The fraction of the total energy radiated via each radiative channel
is nonmonotonic in $\cL$, with the axial channel having a peak at $\cL / \cLcrit \approx 0.75$.
The polar contribution is at least $\approx 85\%$ of the total energy budget for all values of $\cL$
considered.
}
\label{fig:energy_total_fraction}
\end{figure}

Interestingly, the axial and polar contributions to the total energy $\Delta E$ are nonmonotonic with respect to the angular momentum.
In Fig.~\ref{fig:energy_total_fraction}, we show the ratio $\Delta E^{\,(\pm)} / \Delta E$ as a function of
the logarithm of $1 - \cL / \cLcrit$. As we increase the particle's angular momentum (i.e., moving right to left
along the abscissa), we see that $\Delta E^{\,(-)} / \Delta E$ has a local maximum at $\cL \approx 0.75 \, \cLcrit$, yet
with only approximately 14\% of the total energy budget.
We interpret this result as follows. When $\cL = 0$, by symmetry,
all energy must be radiated through the polar channel. As we increase $\cL$, axial perturbations become increasingly excited, and a nonzero (albeit small) percentage of the energy is emitted through them. As $\cL$ approaches $\cLcrit$, the particle orbits the black hole an increasing number of times around the \emph{circular} orbit $r = 4M$. Hence, again by symmetry, we expect the total energy fraction emitted through the polar channel to increase, and it happens to the extent of thereby decreasing the fraction emitted through the axial channel.

\subsection{Time-domain waveforms}

\begin{figure}[t]
\includegraphics{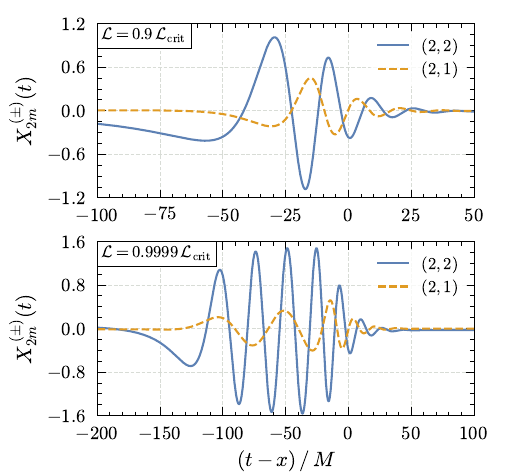}
\caption{Time-domain Zerilli (solid lines) and Regge-Wheeler (dashed lines) mode functions excited by a test particle plunging with $\cL = 0.9 \, \cLcrit$ (top panel) and $\cL = 0.9999 \, \cLcrit$ (bottom panel) into a Schwarzschild black hole. We focus on the dominant quadrupole multipole, associated with the perturbations of each parity: $m=2$ and $m=1$ for the
Zerilli and Regge-Wheeler modes, respectively. The former is always larger in amplitude than the latter.
}
\label{fig:X_td}
\end{figure}

We also computed the time-domain Regge-Wheeler and Zerilli mode functions
using Eq.~\eqref{eq:X_td}. In Fig.~\ref{fig:X_td}, we show the dominant modes
for the Regge-Wheeler and Zerilli functions, $(\ell, m) = (2,1)$ and $(2,2)$,
respectively, as a function of the retarded time $(t - x)/M$. The top panel corresponds
to a particle that plunges with $\cL = 0.9~\cLcrit$, while the bottom panel to $\cL = 0.9999~\cLcrit$.
In the former case, we see that the waveform has the characteristic ``precursor,'' ``sharp burst,'' and ``ringing tail'' phases, as first observed for radial infalling particles by Davis et al.~\cite{Davis:1972ud} and for wave scattering by Vishveshwara~\cite{Vishveshwara:1970zz}.
As $\cL \to \cLcrit$, the large values of $\cL$ cause the Zerilli (but not the Regge-Wheeler) function to have an intermediate quasimonochromatic phase, related to the particle's sweep around $r=4M$. This behavior
is qualitatively the same as that seen in Refs.~\cite{Hadar:2011vj,Folacci:2018cic}
for test particles plunging from the innermost
stable circular orbit $r = 6M$.
We see that the amplitude of the Regge-Wheeler function is always smaller than Zerilli's,
even in the nearest-critical angular-momentum values studied by us.

\begin{figure}[t]
\includegraphics{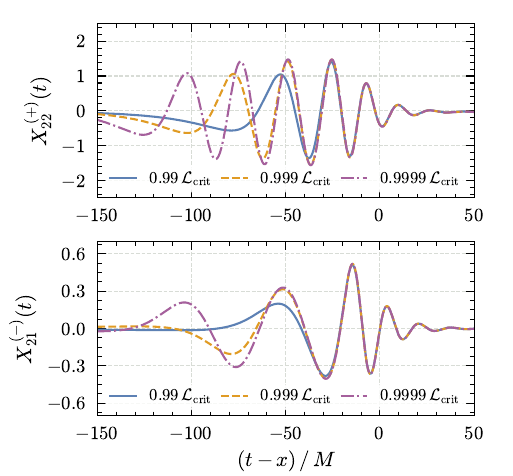}
\caption{Time-domain Zerilli (top panel) and Regge-Wheeler (bottom panel) mode functions excited by a test particle plunging with $\cL = 0.99 \, \cLcrit$ (solid lines), $\cL = 0.999 \, \cLcrit$ (dashed lines) and $\cL = 0.9999 \, \cLcrit$ (dot-dashed lines) into a Schwarzschild black hole.
As in Fig.~\ref{fig:X_td}, we consider the dominant quadrupolar Zerilli and Regge-Wheeler modes.
In the limit $\cL \to \cL_{\rm crit}$, the solutions become quasiuniversal around and after $t - x \gtrsim - 40 \, M$.
}
\label{fig:X_td_uni}
\end{figure}

In Fig.~\ref{fig:X_td_uni}, we take a closer look on the waveforms in the limit $\cL \to \cLcrit$, for the same multipole moments.
In this limit, the waveforms become increasingly similar around their peak amplitudes, $t-x \simeq -40 \, M$, although differences are clearly visible at the ``precursor'' phase.
Similar results hold for the other multipoles we examined.
This result suggests that a quasiuniversal description of the plunge from the marginally stable circular orbit $r = 4M$ exists. Such a treatment for particle's plunging from the innermost
stable circular orbit exists~\cite{Hadar:2009ip}.

\section{The semirelativistic approximation}
\label{sec:kludge}

A complementary way of studying the gravitational wave emission by an infalling particle is via the so-called ``semirelativistic'' approximation~\cite{Ruffini:1981af} (often used in the ``kludge'' approximation~\cite{MorenoGarrido:1995MNRAS,Glampedakis:2002cb,Babak:2006uv}).
In this approach the particle is assumed to move along a fully relativistic geodesic trajectory of the black-hole spacetime while the gravitational wave emission itself is treated approximately, using
the weak-gravity quadrupole formula.

Despite its inherent inconsistency, the semirelativistic approximation is known to perform surprisingly well, when compared against more rigorous results obtained from black-hole perturbation theory.
The price one has to pay for the conceptual and technical simplicity of this
approach is that its accuracy deteriorates as soon as the particle's trajectory enters
the near horizon, strong field regime where radiation backscattering by the spacetime
curvature becomes an important factor.
Unfortunately, this condition is clearly met by a plunging particle so we expect the semirelativistic method to provide accurate results only for the early time portion of the trajectory (i.e.,~the low frequency part of the gravitational wave spectrum.)
Nevertheless, the less accurate
$\omega M \gtrsim 1$ part of the spectrum is of some interest in its own right as it
allows us to understand (and separate) the effects due to the particle's motion and
due to backscattering.

The quadrupole-order gravitational-wave formalism underpinning the semirelativistic approximation can be found in many general relativity textbooks; here we follow and expand the analysis of Ref.~\cite{maggiore2007gravitational}, Sec.~4.3.1, for radially infalling particles.

We start by recalling that the appropriately averaged gravitational-wave
luminosity is given by
\be
L = \tfrac{1}{5} \,  \langle \, \dddot M_{ij} \dddot M^{\,ij} - \tfrac{1}{3} \dddot M ^2 \, \rangle,
\label{eq:power_kludge}
\ee
where the (mass) quadrupole moment $M_{ij}$ is defined as
\be
M^{ij} (t) = \int \dd^3 x \, \rho(t, \vect{x}) \, x^i x^j \,,
\label{eq:quad_mom}
\ee
for a mass density $\rho(t, \vect{x})$, and with trace $M = M^i{}_i$,
which is distinguishable from the black hole's mass $M$ from the context.
The total energy emitted in gravitational waves is given by the integral
\be
\Delta E = \int_{-\infty}^{t_{\rm max}} \dd t \, L (t) \,,
\ee
where the luminosity is to be evaluated without any averaging (also, we
can set $t_{\max} = \infty$ at the particle's horizon crossing time).
The same quantity can be evaluated in the frequency domain
\be
\Delta E =  \int_0^{+\infty} \dd\omega \, \frac{\dd E}{\dd \omega} \,,
\ee
where $ \dd E / \dd \omega $ is the spectral energy distribution.

For a point particle moving along a trajectory $ \vect x_p(t)$ we have
$\rho(t, \vect{x}) = \mu \,\delta^{(3)}[\vect x - \vect x_p(t)]$, and we find
\be
M^{ij} (t) = \mu \, x_p^i(t) \, x_p^j(t) \,.
\ee
As in Sec.~\ref{sec:geo_motion}, the geodesics under consideration can be taken to lie in the equatorial $x$-$y$ plane
and the Cartesian coordinates can be related to the Schwarzschild-Droste coordinates~\eqref{eq:line_element} through Eqs.~\eqref{eq:geo_sch_to_cart}.
In this setup, the only nonvanishing components of the quadrupole moment~\eqref{eq:quad_mom} are $ M_{11}$, $M_{22}$, $M_{12}$, and the trace is $M = M_{11} + M_{12}$.
Then, a short calculation
leads to
\begin{align}
\Delta E &= \frac{2}{15} \int_{-\infty}^{+\infty} \dd t \, [   \dddot{M}_{11}^2 +   \dddot{M}_{22}^2 - \dddot{M}_{11} \dddot{M}_{22} + 3 \dddot{M}_{12}^2 ] \,,
\nonumber
\\
& =  \frac{2}{15 \pi} \int_{0}^{+\infty} \dd\omega \, \omega^6
\, [\,
| \tilde{M}_{11} |^2 +
| \tilde{M}_{22} |^2 +
3 | \tilde{M}_{12} |^2
\nonumber \\
& \quad - \, {\rm Re}[ \tilde{M}_{11}  \tilde{M}^*_{22}] \, ] \,,
\label{Egwtot1}
\end{align}
where the Fourier transform of $M_{ij}$ is defined as\footnote{For clarity, we adopt a slightly different notation for frequency-domain quantities in this section, mirroring Ref.~\cite{maggiore2007gravitational}.}
\be
\tilde M_{ij} (\omega) = \int_{-\infty}^{+\infty} \dd t \,  e^{i \omega t} \, M_{ij}(t) \,.
\label{eq:kludge_FT}
\ee
The fact that $M_{ij}$ is real implies the useful property
\begin{equation}
    \tilde M_{ij}^*(\omega) = \tilde M_{ij}(-\omega) \,.
    \label{eq:quad_real}
\end{equation}

As discussed in Ref.~\cite{maggiore2007gravitational}, the integral~\eqref{eq:kludge_FT} is divergent at
$t \to -\infty$, since $x_p \to +\infty$.
Therefore, some regularization procedure is required.
This is achieved by working
with the Fourier transform $\tilde{\ddot{M}}_{ij}$, which is well behaved at spatial infinity.
In fact, this
procedure is equivalent to the regularization of $\tilde M_{ij} (\omega)$ via integrations by parts, where
the produced divergent boundary terms are discarded (see, e.g., Detweiler and Szedenits~\cite{Detweiler:1979xr} for a
similar regularization of the solution of the Teukolsky equation sourced by a plunging particle). The outcome of this exercise is the regularized quadrupole moment
\begin{equation}
\tilde{M}_{ij}^{\rm reg} =-\tilde{\ddot{M}}_{ij}/\omega^2 \,,
\label{eq:Q_reg}
\end{equation}
and from Eq.~\eqref{Egwtot1} we read off the regularized formula:
\begin{align}
\frac{\dd E}{\dd \omega}  &=
\frac{2 \, \omega^2}{15 \pi} [ \,
| \tilde{\ddot{M}}_{11}  |^2 +
| \tilde{\ddot{M}}_{22} |^2 +
3 | \tilde{\ddot{M}}_{12} |^2
- {\rm Re} [ \tilde{\ddot{M}}_{11} \tilde{\ddot{M}}^*_{22} ] \, ] \,.
\nn
\label{eq:dEdw_kludge}
\end{align}

For the actual numerical evaluation of $\tilde{\ddot M}_{ij}$ (and the subsequent one of $\dd E/ \dd \omega$),
it is advantageous to convert the time integral into a radial integral using the geodesic
equations~\eqref{eq:geo_ode}, leading to
\be
\tilde{\ddot{M}}_{ij} =  \int^{r_{\rm max}}_{2M} \frac{\dd r}{f}
\left [  \frac{2M}{r} + f \, \frac{\cL^2}{r^2}  \right ]^{- \half}  \ddot{M}_{ij} \, e^{i \omega t_p (r)} \,.
\label{eq:quad_final}
\ee
For the problem at hand, the individual $\ddot{M}_{ij}$ components required for the calculation of Eq.~\eqref{eq:Q_reg} are
\begin{subequations}
    \begin{align}
        \ddot{M}_{11}  &= 2 \mu  \left ( x_p \ddot{x}_p + \dot{x}^2_p \right ),
        \\
        \ddot{M}_{22}  &= 2 \mu  \left ( y_p \ddot{y}_p + \dot{y}^2_p \right ),
        \\
        \ddot{M}_{12}  &= \mu \left ( 2 \dot{x}_p \dot{y}_p + x_p \ddot{y}_p + y_p \ddot{x}_p  \right ) .
    \end{align}
\end{subequations}
With the help of Eq.~\eqref{eq:geo_ode}
both the accelerations $\ddot{x}_p$, $\ddot{y}_p$ and the velocities  $\dot{x}_p$, $\dot{y}_p$ can
be rewritten solely as functions of $r$, $\phi_p(r)$ and the orbital constant $\cL$. At the same
time $t_p(r)$ explicitly appears in the exponential term. Both $\phi_p(r)$ and $t_p(r)$ are
obtained numerically as in Sec.~\ref{sec:geo_motion}.
We chose trajectories with $\phi_p (0) = 0$ at some initial radius $r_{\rm max} = r(0)$.
The same radius serves as the ``infinity'' upper limit in the  $\tilde{\ddot M}_{ij}$ integral.
The initial time $t=0$ can be chosen arbitrarily because the addition of a constant in $t_p(r)$ does
not affect the numerical value of Eq.~\eqref{eq:quad_final}.

For any value $\cL < \cL_{\rm crit}$, the integral in the $\tilde{\ddot M}_{11}$ component is the slowest converging one at large $r$ (the absolute value of the integrand decays as $\sim r^{-1/2}$). In order to improve the convergence of this component and reduce the associated numerical error we employ a procedure
similar to that in Sec.~\ref{subsec:general_case_plunge}, namely, we subtract the slowly decaying part prior
to the numerical integration and then add back the (analytically obtained) asymptotic contribution
from that part.

\begin{figure}[t]
\includegraphics{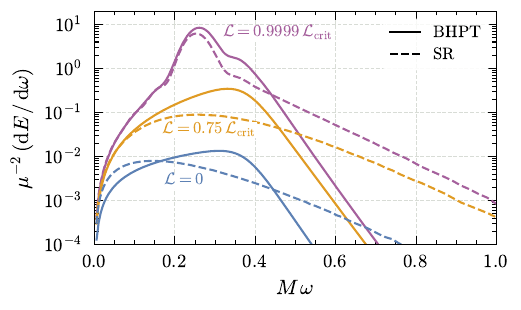}
\caption{The energy spectra for a particle plunging with angular momentum
$\cL = 0$, $\cL = 0.75 \, \cLcrit$, and $\cL = 0.9999 \, \cLcrit$ into
a Schwarzschild black hole in the semirelativistic approximation (``SR,'' dashed lines)
and black-hole perturbation theory, for $\ell = m = 2$ (``BHPT,'' solid lines).
Both calculations agree at low frequencies $M \omega \ll 1$, where the radiation
is due to the particle's quasi-Newtonian motion.
For $\cL = 0$ and $\cL = 0.75 \, \cLcrit$, the kludge calculation
underestimates the location of the spectrum's peak, which is related to the hole's
fundamental quasinormal mode frequency~\cite{Detweiler:1979xr}.
As $\cL \to \cLcrit$, the peak of the spectrum is dominated by the particle's orbit
around $r = 4M$, and the two calculations agree qualitatively with each other up to $M \omega \lesssim 0.25$.
The semirelativistic approximation predicts a slowly decaying tail for the spectrum.
}
\label{fig:kludge_dEdw}
\end{figure}

The semirelativistic energy spectrum, calculated from Eqs.~\eqref{eq:dEdw_kludge} and~\eqref{eq:quad_final}, is shown in Fig.~\ref{fig:kludge_dEdw} and has a characteristic
single-hump profile. When comparing against the full black-hole perturbation theory result for $\ell = m = 2$ shown in Fig.~\ref{fig:dEdw_bhpt},
we notice a moderate disagreement in the location of the emission peak. This is not surprising
because the exact location of this peak depends on the properties of the BH potential near $3M$;
as pointed out earlier, this information is missing altogether in the semirelativistic approximation.
As expected, the two spectra are in good agreement in the low frequency,
$\omega M \ll 1$, end of the spectrum which is the part that corresponds to
quasi-Newtonian motion, that is, when the particle is still moving far
away from the black-hole horizon.
Interestingly, the
agreement improves to some degree as we approach the critical value $\cLcrit$ for scattering.
In this case, the particle spends a large amount of time orbiting  around $r = 4\, M$ and the bulk of
gravitational-wave emission is generated there, as we have discussed in Sec.~\ref{sec:results}.

The opposite high-frequency end of the spectrum, $\omega M \gg 1$, has also some interest of its own.
In the fully relativistic calculation, this part of the spectrum appears to be independent of the angular
momentum $\cL$ and the modal numbers $\ell$ and $m$. We can approximate the high-frequency ``tail'' of $\dd E_{\lm} / \dd \omega$ in the special case of radial infall, taking into account that this part of the spectrum corresponds to the near-horizon region of integration in Eq.~\eqref{eq:out_coef_sch}.
Under these circumstances we can recast the integrand into a single exponential, and expand the exponent
up to linear order in $r / (2 M) - 1$. The integral can then be computed analytically, and it is found
to be dominated by the lower limit of integration. Moreover, in the same high-frequency limit the
Wronskian can be approximated as $W_{\ell m \omega}\simeq 2 i \omega$. These manipulations lead to a
scaling
\begin{equation}
    \frac{\dd E}{\dd \omega} \propto \frac{e^{- 8 \pi M \omega}}{M \omega}
    \quad \textrm{(perturbation theory)} \,.
\end{equation}
The same procedure applied to the semirelativistic spectrum
leads to a slower decaying tail
\begin{equation}
\frac{\dd E}{\dd \omega} \propto \frac{e^{- 4 \pi M \omega}}{M \omega}
\quad \textrm{(semirelativistic)} \,.
\end{equation}
%
The difference can be traced back to the additional highly oscillatory function $X^{(\pm), \, {\rm in}}_{\lm \omega} \simeq \exp(i \omega x)$ in the full black-hole perturbation theory expression, which suppresses the integral at large $\omega$.
This analysis thus explains the difference in the high-frequency tails shown
in Fig.~\ref{fig:kludge_dEdw}.

\section{Conclusions}
\label{sec:conclusions}

We reviewed, and generalized, a transformation by Nakamura, Sasaki, and Shibata~\cite{Shibata:1992zs} that makes the source term of the Zerilli function to have a
faster falloff at spatial infinity, thereby improving the numerical convergence of the convolution integral that arises in the calculation of gravitational radiation by particles in unbound geodesic motion in the Schwarzschild spacetime.

As an application, we studied the gravitational radiation produced by test particles that plunge from rest and with angular momentum $\cL$ from spatial infinity into a Schwarzschild black hole. In particular,
we focused on the limit in which $\cL$ approximates from below the critical value $\cLcrit$
for scattering.
To our knowledge, this is the first time this calculation was done using the
original Regge-Wheeler and Zerilli master functions.
Our results are in agreement with the work by Berti et al.~\cite{Berti:2010ce}
that used the Sasaki-Nakamura equation. We studied in detail the relative contributions
to the energy radiated in gravitational waves due to perturbations of polar and axial parity. We found that the former always dominates. We also observed an quasiuniversal late-time behavior of the
waveforms in limit in which $\cL$ approaches the critical value for scattering, $4M$.

The main merit of the Nakamura-Sasaki-Shibata transformation is that it only requires minimal
modifications to the source term of the Zerilli function [cf.~Eqs.~\eqref{eq:sn_qfun} and~\eqref{eq:source_z_radial_sn}]. The new source term can be easily computed analytically
with any symbolic algebra software.
In contrast, the Sasaki-Nakamura formalism requires the numerical integration of an auxiliary second-order differential equation for the calculation of the source term~\cite{Sasaki:198185,Sasaki:1981kj,Sasaki:1981sx}.
However, an advantage of the Sasaki-Nakamura formalism is that it applies to the Kerr spacetime, while
the method presented here is not.
Our work is alternative to that of Hopper~\cite{Hopper:2017iyq}
that addresses the sources of the Zerilli-Moncrief and Cunningham-Price-Moncrief master functions.

We complemented our calculations in black-hole perturbation theory, with an analysis of the same plunging-particle problem in the semirelativistic approximation.
We found that the two energy spectra agree best with
in the limit $\cL \to \cLcrit$, when
the energy is dominated by the particle's motion around the marginally stable circular orbit.
We also studied the high-frequency limit of the spectrum, expanding upon the discussion in Ref.~\cite{maggiore2007gravitational}.

The method presented here can be used in other problems involving unbound motion
in the Schwarzschild spacetime, including ultrarelativistic plunges~\cite{Ruffini:1973ky,Ferrari:1981dh},
scattering~\cite{Hopper:2017qus}, or in the spacetime of relativistic stars~\cite{Borrelli:1997gk,Ferrari:1999gp,Tominaga:1999iy,Ruoff:2000et} using
the Regge-Wheeler-Zerilli formalism.

\section*{Acknowledgments}
We thank Emanuele Berti, Benjamin Leather, Caio F. B. Macedo,
Raj Patil, Masaru Shibata, Jan Steinhoff, Nicholas C. Stone and Helvi Witek for discussions.
H.O.S acknowledges funding from the Deutsche Forschungsgemeinschaft
(DFG) -- Project No.~386119226.
K.G. acknowledges support from research Grant No.~PID2020-1149GB-I00
of the Spanish Ministerio de Ciencia e Innovaci\'on.
K.Y. acknowledges support from NSF Grants No.~PHY-2207349 and No.~PHY-2309066, a Sloan Foundation Research Fellowship, and the Owens Family Foundation.
This work makes use of the Black Hole Perturbation Toolkit~\cite{BHPToolkit}.
Some of our calculations were performed in the Hypatia cluster at the Max
Planck Institute for Gravitational Physics.

\appendix

\section{The sources of the Regge-Wheeler and Zerilli equations}
\label{app:sources}

In this Appendix we present the sources for the Regge-Wheeler and Zerilli
equations~\eqref{eq:eqs_rwz_td}, quoting from the formulas presented in
Refs.~\cite{Sago:2002fe,Nakano:2003he}; the original derivation is due Zerilli~\cite{Zerilli:1970wzz}.

The source $S^{\ps}_{\lmw}$, which excites the perturbations of polar parity,
is given by
\begin{align}
    S^{\ps}_{\ell m \omega} &= - \, \ii f \ddr{}
    \left[
    \frac{f^2}{\Lambda}
    \left(
    \frac{\ii r}{f} \, \tilde{C}_{1\lmw} + \tilde{C}_{2\lmw}
    \right)
    \right]
    \nn
    &\quad
    + \ii \frac{f^2}{r \Lambda^2}
    \left[
    \ii \frac{\lambda r^2 - 3 \lambda M r - 3 M^2}{r f} \, \tilde{C}_{1\lmw}
    \right.
    \nn
    & \quad \left.
    + \frac{\lambda(\lambda+1)r^2 + 3 \lambda M r + 6 M^2}{r^2} \, \tilde{C}_{2\lmw}
    \right],
    \label{eq:source_z}
\end{align}
where\footnote{We note that in the equation for $\tilde{B}_{\lmw}$, the
term proportional to $A^{(1)}_{\lmw}$ in Eq.~(A42) of Ref.~\cite{Sago:2002fe} has a typo,
which is corrected in Eq.~(4.12) of Ref.~\cite{Nakano:2003he}.}
\begin{align}
\tilde{B}_{\lmw} &=
\frac{8 \pi r^2 f}{\Lambda}
\left[
A_{\lmw} + \frac{1}{\sqrt{\ell (\ell + 1) / 2}} \, B_{\lmw}
\right]
\nn
&\quad -4 \pi \frac{\sqrt{2}}{\Lambda} \frac{M}{\omega} A^{(1)}_{\lmw} \,,
\label{eq:Btilde} \\
\tilde{C}_{1\lmw} &= \frac{8 \pi}{\sqrt{2} \omega} A^{(1)}_{\lmw} + \frac{1}{r} \tilde{B}_{\lmw}
\nn
&\quad
- {16 \pi r} \left[\frac{1}{2}\frac{(\ell + 2)!}{(\ell - 2)!}\right]^{-\tfrac{1}{2}} \, F_{\lmw} \,,
\label{eq:C1tilde} \\
\tilde{C}_{2\lmw} &=
\frac{8 \pi \ii}{\omega \sqrt{\ell(\ell+1)/2}} \frac{r}{f} \, B^{(0)}_{\lmw}
- \frac{\ii}{f} \, \tilde{B}_{\lmw}
\nn
&\quad +
\frac{16 \pi \ii r^2}{f} \left[ \frac{1}{2} \frac{(\ell+2)!}{(\ell-2)!}\right]^{-\tfrac{1}{2}} F_{\lmw} \,.
\end{align}
Here, $f = 1 - 2M/r$, and $A_{\lmw}$, $A^{(1)}_{\lmw}$, $B^{(0)}_{\lmw}$, $B_{\lmw}$, $\tilde{B}_{\lmw}$ and $F_{\lmw}$ are
the Fourier-domain projections of the particle's energy-momentum tensor onto the tensor harmonic basis;
cf.~Ref.~\cite{Sago:2002fe}, Table I, for their expressions in time-domain and generic orbit.
These functions encode information about the particle's geodesic motion, and
when particularized for plunging geodesics they read
\begingroup
\allowdisplaybreaks
\begin{subequations}
\begin{align}
    A_{\lmw} &= \mu \, \frac{V}{r^2 f^2}
    \, Y^{\ast}_{\lm} \, e^{i \omega t_p} \,,
    \\
    A^{(1)}_{\lmw} &= - i \sqrt{2} \, \mu \, \frac{\cE}{r^2 f}
    \, Y^{\ast}_{\lm} \, e^{i \omega t_p} \,,
    \\
    B^{(0)}_{\lmw} &= i \mu \,\frac{\cE \cL}{V r^3} \frac{1}{\sqrt{\ell(\ell+1)/2}}
    %
    \dYdphi \, e^{i \omega t_p} \,,
    \\
    B_{\lmw} &= - \mu \, \frac{\cL}{r^3 f} \frac{1}{\sqrt{\ell(\ell+1)/2}}
    \, \dYdphi \, e^{i \omega t_p} \,,
    \\
    F_{\lmw} &= \mu \,
    \frac{\cL^2}{V r^4}\left[\frac{1}{2}\frac{(\ell + 2)!}{(\ell - 2)!}\right]^{-\tfrac{1}{2}}
    \ddYddphi \, e^{i \omega t_p} ,
\end{align}
\end{subequations}
\endgroup
where, for brevity, we wrote $Y^{\ast}_{\lm} = Y^{\ast}_{\lm}(\pi/2, \phi_p)$ and
defined $V = \sqrt{\cE^2 - U}$, where $U$ is given by Eq.~\eqref{eq:eff_potential}.
Here, $\mu$ is the particle's mass and $t_p$ and $\phi_p$ are functions of $r$, obtained by integrating the geodesic equation~\eqref{eq:geo_ode_radial}.

The source $S^{\mn}_{\lmw}$, that excites the perturbations of axial parity, is given by
\begin{align}
    S^{\mn}_{\lmw} &= \frac{8 \pi i f}{r}
    \left[\frac{1}{2}\frac{(\ell + 2)!}{(\ell - 2)!}\right]^{-\tfrac{1}{2}}
    \left[
        - r^2 \frac{\dd}{\dd r} (f D_{\lmw} )
    \right.
    \nn
    &\quad \left. +\, \sqrt{2 \lambda} \,r f Q_{\lmw} \vphantom{\frac{1}{1}}
    \right],
    \label{eq:source_rw}
\end{align}
where, analogously to the polar case, $D_{\lmw}$ and $Q_{\lmw}$ are the Fourier-domain projections of the particle's energy-momentum tensor
onto the tensor harmonic basis; cf.~Ref.~\cite{Sago:2002fe}, Table I, for their time domain,
general orbit forms.
When particularized for plunging trajectories, $D_{\lmw}$ and $Q_{\lmw}$ become
\begin{subequations}
\begin{align}
    D_{\lmw} &=
    i \mu \, \frac{\cL^2}{V r^4}
    \left[ \frac{1}{2} \frac{(\ell+2)!}{(\ell-2)!} \right]^{-\half}
    \X \, e^{i \omega t_p} \,,
    \\
    Q_{\lmw} &=
    - i \mu \, \frac{\cL}{f r^3}
    \frac{1}{\sqrt{\ell(\ell+1)/2}} \,
    \dYdtheta \, e^{i \omega t_p} \,,
\end{align}
\end{subequations}
where we introduced the shorthand notation,
\begin{subequations}
\begin{align}
    X_{\lm} = 2 \pd_{\phi} ( \pd_{\theta} - \cot\theta ) Y_{\lm} \,.
\end{align}
\end{subequations}
In the case of radial infall, $\cL = 0$, both $D_{\lmw}$ and $Q_{\lmw}$ are
zero, and consequently $S^{\mn}_{\lmw}$, vanishes.

\section{The near-horizon and far-field expansions of the Regge-Wheeler
and Zerilli functions}
\label{app:up_sol}

When integrating numerically the homogeneous Regge-Wheeler and Zerilli equations,
we use series representations of the solutions to these equations in the
near-horizon [$r/(2M) - 1 \ll 1$ or $x \to -\infty$] and asymptotic ($r \gg 2M$ or $x \to \infty$) regions
of the Schwarzschild spacetime.
The coefficients in these series satisfy certain recursion relations that we
summarize here.

We first consider the near-horizon limit. We assume that the Regge-Wheeler and Zerilli mode
functions $\mode$ can be factorized as,
\begin{equation}
    \mode =  e^{- i \omega x} \sum_{n=0}^{\infty} c_n^{\,(\pm)} \, z(r)^n \, \,,
    \quad z = r/(2M) - 1\,.
    \label{eq:mode_nh_exp}
\end{equation}
We substitute Eq.~\eqref{eq:mode_asympt_exp} in the homogeneous
equation~\eqref{eq:eqs_rwz_hom} and derive a recursion relation between the
coefficients $c_n^{\,(\pm)}$.
For the Zerilli equation this relation is
\begin{align}
\mathfrak{a} \, c^{\,(+)}_n &=
\mathfrak{b} \, c_{n-1}^{\,(+)}
+ \mathfrak{c} \, c_{n-2}^{\,(+)}
+ \mathfrak{d} \, c_{n-3}^{\,(+)}
+ \mathfrak{e} \, c_{n-4}^{\,(+)}
+ \mathfrak{f} \, c_{n-5}^{\,(+)} \,,
\nn
\label{eq:rec_z_nh}
\end{align}
where
\begin{align}
\mathfrak{a} &= n (3 + 2 \lambda)^2 (n - 4 i \sigma) \,,
\nn
\mathfrak{b} &= (3+2\lambda) \, \{ 4 \lambda ^2 + i (40 \lambda + 36) (n-1) \sigma
\nn
             &\quad + \lambda  [2 (9-4 n) n-6] - 3 (n-2) (2 n-1) \} \,,
\nn
\mathfrak{c} &= 24 \lambda ^3 - \lambda ^2 (24 n^2 - 108 n + 72)
- \lambda ( 36 n^2 - 168 n + 174)
\nn
&\quad -9 n^2 + i (160 \lambda ^2+288 \lambda +108) (n-2) \sigma +45 n-54 \,,
\nn
\mathfrak{d} &= 24 \lambda ^3 - \lambda ^2 (16 n^2 - 108 n + 144)
- \lambda  (12 n^2 - 84 n + 144)
\nn
&\quad + i (160 \lambda ^2+192 \lambda +36) (n-3) \sigma \,,
\nn
\mathfrak{e} &= i \lambda \left(80 \lambda + 48 \right) (n-4) \sigma -4 \lambda ^2 (n-6) (n-3) + 8 \lambda^3 \,,
\nn
\mathfrak{f} &= 16 i \lambda ^2 (n-5) \sigma \,,
\end{align}
and $\sigma = M \omega$~\cite{Chandrasekhar:1985kt}.
For the Regge-Wheeler equation we obtain instead
\begin{align}
n(n + 4i\sigma) \, c_{n}^{\,(-)} &=
- [ 2 n^2 - 12 i (n-1) \sigma - 5 n + \ell(\ell+ 1)
\nn
&\quad -6 ] \, c_{n-1}^{\,(-)}
- [(n-\ell -3) (n+\ell -2)
\nn
&\quad - 12 i (n-2) \sigma] \, c_{n-2}^{\,(-)}
+ 4 i (n-3) \sigma \, c_{n-3}^{\,(-)}.
\nn
\label{eq:rec_rw_nh}
\end{align}
In Eqs.~\eqref{eq:rec_z_nh} and~\eqref{eq:rec_rw_nh}, $c^{\,(\pm)}_{n} = 0$ for negative
values of $n$.

We now consider the asymptotic expansion of $\mode$. We assume they can be factorized as,
\begin{equation}
    \mode = J^{\,(\pm)}_{\lmw}(r) \, e^{+ i \omega x} \,,
    \label{eq:mode_asympt_exp}
\end{equation}
where $J^{\,(\pm)}_{\lmw}$ (``Jost function'') approaches a constant as $x \to \infty$ in order to
recover a purely outgoing behavior.
We then substitute Eq.~\eqref{eq:mode_asympt_exp} in the homogeneous
equation~\eqref{eq:eqs_rwz_hom}. This results in a differential equation
for $J^{\,(\pm)}_{\lmw}$,
\begin{equation}
    \left[
        f \frac{\dd^2}{\dd r^2}
        + \left(\frac{2M}{r^2} + 2 i \omega \right) \frac{\dd}{\dd r}
        - \frac{V^{\,(\pm)}_{\ell}}{f}
    \right]J^{\,(\pm)}_{\lmw} = 0
    \,,
\label{eq:just_equation}
\end{equation}
which we solve with Frobenius' method. We assume a series expansion
of the form,
\begin{equation}
    J^{\,(\pm)}_{\lmw} = \sum_{n = 0}^{\infty} {a^{\,(\pm)}_{n}} / (\omega r)^{n} \,.
\end{equation}
We then substitute this expansion in Eq.~\eqref{eq:just_equation}, and
derive a recursion relation between the coefficients $a^{\,(\pm)}_n$.
When using the Zerilli potential~\eqref{eq:pot_zerilli}, we obtain
%
\begin{align}
    \label{eq:rec_z_coefs}
    2 i \lambda^2 n a^{\,(+)}_{n} &=
    \lambda [\lambda (n - 1) n - 12 i \sigma (n - 1) - 2 \lambda ( \lambda + 1 ) ] \, a^{\,(+)}_{n-1}
    \nn
    &\quad + 2 \sigma [ \lambda (3 - \lambda) (n-2) (n-1) - (\lambda^2 + 9 i \sigma)
    \nn
    &\quad \times (n - 2) - 3 \lambda^2] \, a^{\,(+)}_{n-2}
    + 3 \sigma^2 [ (3 - 4 \lambda) (n - 3) ]
    \nn
    &\quad \times (n-2) - 4 \lambda (n-3) - 6 \lambda] \, a^{\,(+)}_{n-3}
    \nn
    &\quad - 18 \sigma^3 (n - 3)^2 \, a^{\,(+)}_{n-4}
    \,,
\end{align}
\\
while using the Regge-Wheeler potential~\eqref{eq:pot_regge_wheeler}
we find
%
\begin{align}
    \label{eq:rec_rw_coefs}
    2 i n a^{\,(-)}_{n} &=
    - 2 \sigma [(n+1) (n-3)]  \, a^{\,(-)}_{n-1}
    \nn
    &\quad - [\ell(\ell+1) - n(n-1)] \, a^{\,(-)}_{n-2}
    \,,
\end{align}
where $a^{\,(\pm)}_n = 0$ for negative values of $n$~\cite{Chandrasekhar:1975zza,Hopper:2010uv}.
The derivations we just made also apply to modes that behave as $\simeq \exp(- i \omega x)$ at
spatial infinity. In practice, we can replace $\sigma \to - \sigma$ in Eqs.~\eqref{eq:rec_z_coefs} and~\eqref{eq:rec_rw_coefs}.

\section{Calculation of the Wronskian}
\label{app:how_to_find_wronskian}

To calculate the Wronskian~\eqref{eq:wronskian_infty}, we need first to calculate the mode amplitude
$A_{\lmw}^{\,(\pm)\, {\rm in}}$.
We do this calculation following the same strategy outlined in Ref.~\cite{Berti:2010ce}, Appendix A1,
in the context of the Sasaki-Nakamura formalism.

We first integrate $\modeIn$ from near the horizon [making use of the recursion relations~\eqref{eq:rec_z_nh} and~\eqref{eq:rec_rw_nh}] out to some large $r_{\rm max}$.
This gives us two constants
$\modeIn(r_{\rm max})$ and $\dd \modeIn(r_{\rm max}) / \dd r$.
%
%
For large $r$, the mode function is approximately given by
\begin{align}
\modeIn &\simeq
A^{\,(\pm),\, {\rm in}}_{\lmw} \, e^{-i \omega x}
+ A^{\,(\pm),\, {\rm out}}_{\lmw} \, e^{+i \omega x}
\nn
&=
A^{\,(\pm),\, {\rm in}}_{\lmw} \, J^{\,(\pm),\,{\rm in}}_{\lmw} \, e^{-i \omega x}
+ A^{\,(\pm),\, {\rm out}}_{\lmw} \, J^{\,(\pm),\,{\rm out}}_{\lmw} \, e^{+i \omega x},
\nn
\label{eq:modeIn_jost}
\end{align}
where we used the factorization~\eqref{eq:mode_asympt_exp} in the second line.

We evaluate the Jost functions $J_{\lmw}^{\,(\pm)}$ using the recursion relations~\eqref{eq:rec_z_coefs} and~\eqref{eq:rec_rw_coefs}, with $a_0^{\,(\pm)} = 1$ and
adding terms in the series until a sufficient level of accuracy is reached.
Equation~\eqref{eq:modeIn_jost} and its derivative with respect to $r$ provide two equations that depend on $\modeIn$ (and its derivative) at $r_{\rm max}$; the outcomes of the numerical integration of $\modeIn$.
We use these two equations to solve for the two amplitudes $A_{\lmw}^{\,(\pm),\, {\rm in}}$ and
$A_{\lmw}^{\,(\pm),\, {\rm out}}$. The former is then used to calculate the Wronskian~\eqref{eq:wronskian_infty}.

\bibliography{biblio}
\end{document}